\documentclass[onecolumn,showpacs,preprintnumbers,elsart]{revtex4}
\usepackage{makeidx}
\usepackage{amssymb}
\usepackage{amsmath}
\usepackage{mathrsfs}
\usepackage{graphicx}
\usepackage{dcolumn}
\usepackage{bm}
\usepackage[center]{subfigure}
\usepackage{color}

\begin{document}

\title{Two-dimensional dipolar gap solitons in free space with spin-orbit
coupling}
\author{Yongyao Li$^{1,2}$, Yan Liu$^{2}$, Zhiwei Fan$^{2}$, Wei Pang$^{3}$,
Shenhe Fu$^{4}$}
\email{fushenhe@jnu.edu.cn}
\author{Boris A. Malomed$^{5,6,1}$}
\affiliation{$^{1}$School of Physics and Optoelectronic Engineering, Foshan University,
Foshan 528000, China \\
$^{2}$College of Electronic Engineering, South China Agricultural
University, Guangzhou 510642, China \\
$^{3}$ Department of Experiment Teaching, Guangdong University of
Technology, Guangzhou 510006, China\\
$^{4}$Department of Optoelectronic Engineering, Jinan University, Guangzhou
510632, China\\
$^{5}$ Department of Physical Electronics, School of Electrical Engineering,
Faculty of Engineering, Tel Aviv University, Tel Aviv 69978, Israel.\\
$^{6}$Laboratory of Nonlinear-Optical Informatics, ITMO University, St.
Petersburg 197101, Russia}

\begin{abstract}
We present gap solitons (GSs) that can be created in free nearly
two-dimensional (2D) space in dipolar spinor Bose-Einstein condensates with
the spin-orbit coupling (SOC), subject to tight confinement, with size $%
a_{\perp }$, in the third direction. For quasi-2D patterns, with lateral
sizes $l\gg a_{\perp }$, the kinetic-energy terms in the respective spinor
Gross-Pitaevskii equations may be neglected in comparison with SOC. This
gives rise to a bandgap in the system's spectrum, in the presence of the
Zeeman splitting between the spinor components. While the present system
with contact interactions does not produce 2D solitons, stable gap solitons
(GSs), with vorticities $0$ and $1$ in the two components, are found, in
quasi-analytical and numerical forms, under the action of dipole-dipole
interaction (DDI). Namely, isotropic and anisotropic 2D GSs are obtained
when the dipoles are polarized, respectively, perpendicular or parallel to
the 2D plane. The GS families extend, as \textit{embedded solitons} (ESs),
into spectral bands, a part of the ES branch being stable for isotropic
solitons. The GSs remain stable if the competing contact interaction, with
the sign opposite to that of the DDI, is included, while the addition of the
contact term with the same sign destabilizes the GSs, at first replacing
them by breathers, and eventually leading to destruction of the solitons.
Mobility and collision of the GSs are studied too, revealing negative and
positive effective masses of the isotropic and anisotropic solitons,
respectively.
\end{abstract}

\maketitle

\section{Introduction}

Gap solitons (GSs) are usually defined as self-trapped modes existing in
spectral bandgaps of periodic potentials. GSs have been predicted and
observed in diverse optical media, such as Bragg gratings \cite{Krug},
waveguide arrays \cite{arrays}, and photonic crystals \cite{Segev} (see also
reviews \cite{DS-review0,DS-review1,DS-review2}), as well as in
Bose-Einstein condensates (BECs) trapped in optical lattices \cite%
{BEC,GS-BEC}, and in a plasmonic medium including a lattice potential \cite%
{plasmon}. The dynamics of spatial GSs is usually modeled by the nonlinear
Schr\"{o}dinger/Gross-Pitaevskii equations (NLSEs/GPEs) with periodic
potentials. GSs in fiber gratings are described by nonlinear coupled-mode
equations for counterpropagating waves \cite{DS-review0}. In all cases, the
underlying spatially periodic structures play a key role, producing spectral
bandgaps in which GSs can be created.

A challenging problem in optics and BEC is the creation of stable
fundamental and vortical bright solitons in two- and three-dimensional (2D
and 3D) geometry \cite{old-review,Dumitru,recent-review}. New possibilities
are offered by spin-orbit coupling (SOC) in spinor BEC \cite{SOC1} and its
counterparts in optics \cite{Barcelona,Sakaguchi2}. In particular, the
interplay of the linear SOC with the cubic attractive nonlinearity opens a
way for creating 2D ground-state \cite{Sakaguchi,Barcelona} and 3D
metastable \cite{HP} solitons in \emph{free space}, which was previously
deemed impossible (1D \cite{Kartashov1D} and 2D \cite{Kartashov2D} GSs
supported by a combination of lattice potentials and SOC were predicted too).

We aim to demonstrate that SOC offers another unexpected possibility, to
create stable 2D GSs in free space, without the use of any periodic
potential. The model is formulated in Section II, where we consider the 2D
condensate under the action of strong SOC, which makes it possible to
neglect the kinetic-energy terms in the corresponding system of coupled
GPEs, thus reducing them to a first-order system of coupled-mode equations.
The gap in the system's spectrum is generated by the Zeeman splitting (ZS),
which is an essential ingredient of SOC settings \cite{Zeeman}. However,
numerical results demonstrate that the usual contact nonlinearity of any
sign fails to build solitons in this 2D system. Our main result is
prediction of families of stable isotropic and anisotropic 2D\ GSs under the
action of dipole-dipole interaction (DDI). We note that the realization of
SOC in the condensate of dipolar chromium atoms was proposed in Ref. \cite%
{Cr}, and elaborated theoretically in subsequent works \cite{Xunda}.

Using a combination of an analytical approximation, which is available close
to edges of the bandgap (similar to the approximation recently developed in
Ref. \cite{Sherman}), and numerical methods, in Section III we construct
families of stationary solutions for isotropic and anisotropic GSs in the
condensates composed of dipoles oriented, respectively, perpendicular or
parallel to the system's plane. The isotropic and anisotropic families
extend, severally, as \textit{embedded solitons} (ESs) \cite{Jianke1999}
across the top and bottom edge of the bandgap into the adjacent Bloch band.
Stability of the GSs and ESs is explored by means of systematic direct
simulations. Mobility and collisions of the GSs are studied in Section IV,
showing that the effective mass is positive for the anisotropic 2D solitons,
and \textit{negative} for the isotropic ones (negative mass is a known
property of GSs \cite{negative-mass}). In Section V, we consider an extended
system, which includes both the dipole-dipole and contact interactions. The
conclusion is that the solitons persist as stable modes under the action of
\textit{competing} interactions with opposite signs. If the signs are
identical, the addition of the contact interaction tends to destabilize a
soliton, replacing it by a breather. Eventually, all self-trapped modes are
destroyed. The paper is concluded by Section VI.

\section{The model}

In the scaled form (the corresponding estimates of physical parameters are
given below), the coupled GPEs for two components of the spinor wave
function, $\phi _{\pm }$, carrying the same magnetic moment and coupled by
the SOC of the Rashba type \cite{Rashba}, are \cite%
{SOCsoliton1,SOCsoliton2,Sakaguchi,Sherman,Xunda,Guihua2017}:%
\begin{gather*}
i{{\partial _{t}}}\phi _{+}=-\left( 2m\right) ^{-1}\nabla ^{2}\phi
_{+}+\lambda \left( {{\partial _{x}}}-i{{\partial _{y}}}\right) \phi
_{-}-\Omega \phi _{+} \\
+\left( g|\phi _{+}|^{2}+\tilde{g}|\phi _{-}|^{2}\right) \phi _{+}+\kappa
\phi _{+}\int R(\mathbf{r}-\mathbf{r^{\prime }})(|\phi _{+}(\mathbf{%
r^{\prime }})|^{2}+|\phi _{-}(\mathbf{r^{\prime }})|^{2})d\mathbf{r^{\prime }%
},
\end{gather*}%
\begin{gather}
i{{\partial _{t}}}\phi _{-}=-\left( 2m\right) ^{-1}\nabla ^{2}\phi _{\pm
}-\lambda \left( {{\partial _{x}}}+i{{\partial _{y}}}\right) \phi
_{+}+\Omega \phi _{-}  \notag \\
+\left( g|\phi _{-}|^{2}+\tilde{g}|\phi _{+}|^{2}\right) \phi _{-}+\kappa
\phi _{-}\int R(\mathbf{r}-\mathbf{r^{\prime }})(|\phi _{+}(\mathbf{%
r^{\prime }})|^{2}+|\phi _{-}(\mathbf{r^{\prime }})|^{2})d\mathbf{r^{\prime }%
},  \label{Fulleq}
\end{gather}%
where $m$ is the atomic mass, {$g$ and }$\tilde{g}$ are{\ strengths of the
contact self- and cross-interactions (usually, these strengths are nearly
equal in mixtures of different states of the same atomic species \cite{Ho}),}
while $\lambda $, $\Omega >0$, and $\kappa >0$ represent SOC, ZS, and DDI,
respectively \cite{Cr}. Note that the sign of $\kappa $ may be altered by
means of a rotating magnetic field \cite{Giovanazzi}, and ZS may be replaced
by the Stark - Lo Surdo splitting in dc electric field.

\begin{figure}[tbp]
\centering{\label{fig1a} \includegraphics[scale=0.3]{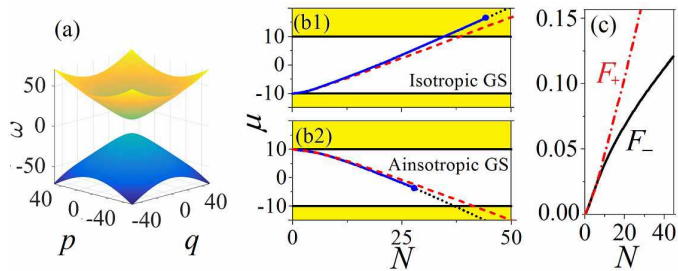}}
\caption{(a) The bandgap structure, as per Eq. (\protect\ref{linear}) with $(%
\protect\lambda ,\Omega)=(1,10)$. The same values of $\Omega $ and $\protect%
\lambda $ are used below throughout the paper. (b) The chemical potential of
the isotropic (b1) and anisotropic (b2) 2D solitons, vs. their total norm, $%
N $, defined as per Eq. (\protect\ref{N}). Yellow areas are spectral bands,
with the white gap between them. Blue solid and black dotted curves are,
respectively, numerically found stable branches and their unstable
extensions. The stable branch of isotropic solitons [semi-vortices, see Eq. (%
\protect\ref{SV})] traverses the bandgap and extends, in the form of ESs, to
the upper band. It is stable at $N<N_{\max }\approx 45.0$. The branch of
anisotropic solitons extends into the lower band, but its stable segment
terminates in the gap, at $N=N_{\max }\approx 28.2$. Red dashed curves are
semi-analytical predictions based on Eq. (\protect\ref{isotropic}). (c) The
share of the total norm in the vortex component, $F_{\mp }=N_{\mp }/N$, vs. $%
N$ for the isotropic ($-$) and anisotropic ($+$) solitons. Note that figure
and all others refer to the scaled norm, while the actual number of atoms in
the respective BECis $N_{\mathrm{at}}\sim 10^{3}N$, see Eq. (\protect\ref%
{Nat}). In all the panels, we have fixed $\protect\kappa=0.1$.}
\label{Spectra}
\end{figure}

SOC of the Rashba type is adopted here as it helps to build 2D solitons,
while SOC terms of the Dresselhaus type tend to cause delocalization \cite%
{Sherman}. Nevertheless, Eqs. (\ref{isotropic}) and (\ref{time}), derived
below in the limit case when one component is much larger than the other,
take a universal form for any combination of the Rashba and Dresselhaus
terms. Although, strictly speaking, those asymptotic equations are valid
only in small vicinities of the bottom and top edges of the spectral
bandgap, see Eq. (\ref{close}) below, the results reported in the next
section [see Fig. \ref{Spectra}(b)] clearly demonstrate that the predictions
produced by Eqs. (\ref{isotropic}) and (\ref{time}) are quite accurate in
the entire bandgap, hence the universality implied by those equations
remains approximately valid for the full GS families.

In the isotropic setting, with dipoles oriented perpendicular to the $\left(
x,y\right) $ plane, the respective repulsive DDI kernel is
\begin{equation}
R_{\mathrm{iso}}(\mathbf{r}-\mathbf{r^{\prime }})={1/[\epsilon ^{2}+(\mathbf{%
r}-\mathbf{r^{\prime }})^{2}]^{3/2}},  \label{iso}
\end{equation}
where cutoff $\epsilon $ is determined by the confinement in the transverse
dimension \cite{Santos,epsilon,quadr}, while effects of the attractive DDI
in the transverse direction are suppressed by the tight confinement, whose
strength (trapping frequency) is much larger than chemical potential
produced by the effective 2D GPE \cite{Tikhonenkov}. If the dipoles are
polarized parallel to the $\left( x,y\right) $ plane \cite{Tikhonenkov}, the
DDI is anisotropic, with
\begin{equation}
R_{\mathrm{aniso}}(\mathbf{r}-\mathbf{r^{\prime }})={(1-3\cos ^{2}\Theta
)/[\epsilon ^{2}+(\mathbf{r}-\mathbf{r^{\prime }})^{2}]^{3/2}},
\label{aniso}
\end{equation}
where $\Theta $ is the angle between the polarization direction and ${(%
\mathbf{r}-\mathbf{r^{\prime }})}$. The simplest approximation for the
cutoff, adopted here, is sufficient, as the detailed analysis demonstrates
that the exact form gives rise to practically the same result \cite%
{Santos,epsilon,quadr}.

The SOC\ coefficient relevant to experimental settings with
transverse-confinement size $a_{\perp }$ is estimated, in physical units, as
$\lambda \gtrsim \hbar ^{2}/\left( ma_{\perp }\right) $ \cite{Drummond-Pu}.
It follows from here that the kinetic-energy terms in Eq. (\ref{Fulleq}) may
be neglected for all quasi-2D patterns, with lateral sizes $l\gg a_{\perp }$%
, reducing it to the coupled-mode equations,%
\begin{gather}
i{{\partial _{t}}}\phi _{+}=\lambda \left( {\partial _{x}}-i{{\partial _{y}}}%
\right) \phi _{-}-\Omega \phi _{+}+\left( g|\phi _{+}|^{2}+\tilde{g}|\phi
_{-}|^{2}\right) \phi _{+}  \notag \\
+\kappa \phi _{+}\int R(\mathbf{r}-\mathbf{r^{\prime }})\left[ |\phi _{+}(%
\mathbf{r^{\prime }})|^{2}+|\phi _{-}(\mathbf{r^{\prime }})|^{2}\right] d%
\mathbf{r^{\prime },}  \label{Basiceq01}
\end{gather}%
\begin{gather}
i{{\partial _{t}}}\phi _{-}=-\lambda \left( {\partial _{x}}+i{{\partial _{y}}%
}\right) \phi _{+}+\Omega \phi _{-}+\left( g|\phi _{-}|^{2}+\tilde{g}|\phi
_{+}|^{2}\right) \phi _{-}  \notag \\
+\kappa \phi _{-}\int R(\mathbf{r}-\mathbf{r^{\prime }})\left[ |\phi _{+}(%
\mathbf{r^{\prime }})|^{2}+|\phi _{-}(\mathbf{r^{\prime }})|^{2}\right] d%
\mathbf{r^{\prime },}  \label{Basiceq02}
\end{gather}%
cf. Ref. \cite{Sakaguchi2} in 1D. In a different context, a 2D SOC system
with a deep optical-lattice potential and negligible kinetic energy was
recently introduced in Ref. \cite{flatband}.

The linear part of Eqs. (\ref{Basiceq01}) and (\ref{Basiceq02}) is
essentially the same as in the 2D nonlinear Dirac equation (NDE) for bosons
in honeycomb lattices \cite{Haddad1}, which gives rise to stable states in
the presence of an external trap \cite{Haddad2}. In the absence of the trap,
stable 2D solitons were produced by NDE with the \emph{sign-indefinite}
contact nonlinear terms, $\pm (|\phi _{+}|^{2}-|\phi _{-}|^{2})\phi _{\pm }$%
, in its two components \cite{Kevrekidis}. However, this case is not
relevant to the BEC\ spinor wave function, whose components represent atomic
states with nearly equal intra- and inter-component scattering lengths \cite%
{Ho}. We have checked that Eqs. (\ref{Basiceq01}) and (\ref{Basiceq02}) with
the corresponding contact nonlinearity, $\sim \left( |\phi _{+}|^{2}+|\phi
_{-}|^{2}\right) \phi _{\pm }$, fail to create 2D solitons, unlike the
system which includes the kinetic energy \cite{Sakaguchi}. This is explained
by the fact that the gradient energy corresponding to the Dirac operator in
Eqs. (\ref{Basiceq01}) and (\ref{Basiceq02}), unlike its Schr\"{o}dinger
counterpart, cannot balance the local cubic nonlinearity. However, it is
shown below that GSs are readily produced by the balance of the SOC with the
DDI, as well as with a combination of the DDI and contact interaction with
opposite signs. The statements concerning the balance are corroborated by
analytical results reported below for GSs located close to bandgap edges.

Linearizing Eqs. (\ref{Basiceq01}) and (\ref{Basiceq02}) for $\phi _{\pm
}\sim \exp \left( ipx+iqy-i\omega t\right) $, we derive the dispersion
relation,
\begin{equation}
\omega ^{2}=\Omega ^{2}+\lambda ^{2}(p^{2}+q^{2}),  \label{linear}
\end{equation}%
with the \emph{free-space} bandgap of width $2\Omega $, as shown in Fig. \ref%
{Spectra}(a) for $\lambda =1$, $\Omega =10$ (by means of scaling, these
values of the ZS and SOC strengths are fixed in the present work). The
bandgap will close if the small kinetic-energy terms are kept in Eq. (\ref%
{Fulleq}), which may cause transformation of the GSs into usual solitons at
times much larger than experimentally relevant time scales.

GS solutions to Eqs. (\ref{Basiceq01}) and (\ref{Basiceq02}) with chemical
potential $\mu $ are looked for as $\phi _{\pm }=e^{-i\mu t}u_{\pm }(x,y)$,
with $u_{\pm }$ obeying equations%
\begin{gather}
\mu u_{+}=\lambda \left( {{\partial _{x}}}-i{{\partial _{y}}}\right)
u_{-}-\Omega u_{+}+\left( g|\phi _{+}|^{2}+\tilde{g}|\phi _{-}|^{2}\right)
u_{+}  \notag \\
+\kappa u_{+}\int R(\mathbf{r}-\mathbf{r^{\prime }})\left[ |u_{+}(\mathbf{%
r^{\prime }})|^{2}+|u_{-}(\mathbf{r^{\prime }})|^{2}\right] d\mathbf{%
r^{\prime }}.  \label{Basiceq1}
\end{gather}%
\begin{gather}
\mu u_{-}=-\lambda \left( {{\partial _{x}}}+i{{\partial _{y}}}\right)
u_{+}+\Omega u_{-}+\left( g|\phi _{-}|^{2}+\tilde{g}|\phi _{+}|^{2}\right)
u_{-}  \notag \\
+\kappa u_{-}\int R(\mathbf{r}-\mathbf{r^{\prime }})\left[ |u_{+}(\mathbf{%
r^{\prime }})|^{2}+|u_{-}(\mathbf{r^{\prime }})|^{2}\right] d\mathbf{%
r^{\prime }}.  \label{Basiceq2}
\end{gather}%
The GS is characterized by the norm, which is proportional to the total
number of atoms in the binary BEC,%
\begin{equation}
N=N_{+}+N_{-}\equiv \int \left( |u_{+}(\mathbf{r})|^{2}+|u_{-}(\mathbf{r}%
)|^{2}\right) d\mathbf{r}.  \label{N}
\end{equation}

Furthermore, Eqs. (\ref{Basiceq1}) and (\ref{Basiceq2}) with the isotropic
DDI kernel given by Eq. (\ref{iso}) admit solutions with an \emph{exact
structure} of \textit{semi-vortices} \cite{Sakaguchi,HP,Sherman}, i.e.,
isotropic complexes with zero vorticity $S=0$ in component $u_{+}$, and
vorticity $S=1$ in $u$:%
\begin{equation}
u_{+}=U_{+}(r),~u_{-}=U_{-}(r)e^{i\theta },  \label{SV}
\end{equation}%
where $\left( r,\theta \right) $ are the polar coordinates in the $\left(
x,y\right) $ plane, and radial functions $U_{\pm }(r)$ are real. Indeed, the
substitution of the semi-vortex ansatz (\ref{SV}) in Eqs. (\ref{Basiceq1})
and (\ref{Basiceq2}) yields, after performing the angular integration in the
DDI terms, to the following equations which include solely the radial
coordinate:%
\begin{gather}
\mu U_{+}=\lambda \left( \frac{dU_{-}}{dr}+\frac{1}{r}U_{-}\right) -\Omega
U_{+}+\left( gU_{+}^{2}+\tilde{g}U_{-}^{2}\right) U_{+}  \notag \\
+\kappa U_{+}\int_{0}^{\infty }\varrho (r,r^{\prime })\left[ U_{+}^{2}(r%
\mathbf{^{\prime }})+U_{-}^{2}(r\mathbf{^{\prime }})\right] r^{\prime }dr%
\mathbf{^{\prime }}.  \label{radial1}
\end{gather}%
\begin{gather}
\mu U_{-}=-\lambda \frac{dU_{+}}{dr}+\Omega U_{-}+\left( gU_{-}^{2}+\tilde{g}%
U_{+}^{2}\right) U_{-}  \notag \\
+\kappa U_{-}\int \varrho (r,r^{\prime })\left[ U_{+}^{2}(r\mathbf{^{\prime }%
})+U^{2}_{-}(r\mathbf{^{\prime }})\right] r^{\prime }dr\mathbf{^{\prime }},
\label{radial2}
\end{gather}%
where the effective radial kernel is
\begin{eqnarray}
\varrho (r,r^{\prime }) &=&\frac{2E(k)}{\sqrt{\epsilon ^{2}+\left(
r+r^{\prime }\right) ^{2}}\left[ \epsilon ^{2}+\left( r-r^{\prime }\right)
^{2}\right] }.  \notag \\
k &\equiv &\frac{2\sqrt{rr^{\prime }}}{\sqrt{\epsilon ^{2}+\left(
r+r^{\prime }\right) ^{2}}}.  \label{rho}
\end{eqnarray}%
Here, $E(k)$ is the standard complete elliptic integral of the second kind
with modulus $k$.

Numerical results presented in the next section, see Figs. \ref{comparison}%
(a,b,c) and \ref{solitonsolution}(a1,a2,a3), confirm that Eqs. (\ref%
{Basiceq1}) and (\ref{Basiceq2}) with the isotropic kernel indeed give rise
to the two-component solitons whose structure precisely conforms to ansatz (%
\ref{SV}). Furthermore, it is demonstrated below that the anisotropic kernel
gives rise, as a matter of fact, to deformed patterns of the same
semi-vortex type, with $S=0$ in component $u_{-}$ and $S=-1$ in $u_{+}$, see
Figs. \ref{comparison}(d,e,f) and \ref{solitonsolution}(b1,b2,b3) below.

\section{Families of gap solitons and embedded solitons}

\subsection{Isotropic and anisotropic solitons near edges of the bandgap}

\begin{figure}[tbp]
\centering{\label{fig1b} \includegraphics[scale=0.4]{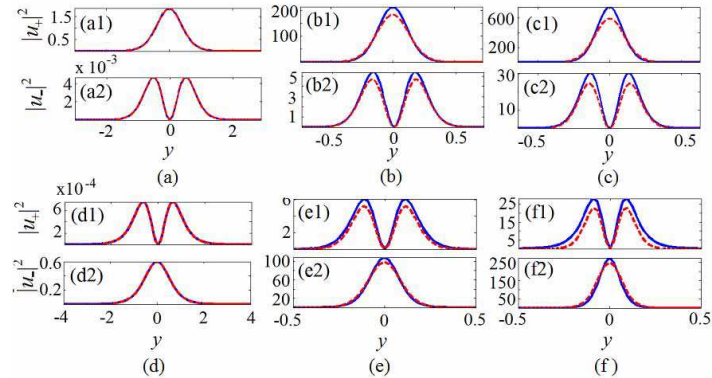}}
\caption{Comparison between cross sections, drawn along $x=0$, of stable
isotropic solitons [semi-vortices, see Eq. (\protect\ref{SV})] (a,b,c) and
anisotropic (d,e,f) solitons, as found in the numerical form (blue solid
curves), and predicted by Eqs. (\protect\ref{un_bottom}) and (\protect\ref%
{isotropic}) (red dash curves). (a) A GS with $N=2.013$ and $\protect\mu %
=-9.7$. (b) A GS with $N=21.37$ and $\protect\mu =1.72$. (c) An ES with $%
N=44.47$, $\protect\mu =16.65$, located close to the stability boundary (the
blue dot) of isotropic solitons. (d) A GS with $N=2.0043$, $\protect\mu %
=9.935$. (e) A GS with $N=16.12$, $\protect\mu =4.06$. (f) A GS with $%
N=28.17 $, $\protect\mu =-3.78$, located close to the stability boundary of
anisotropic solitons.}
\label{comparison}
\end{figure}

Close to edges of the bandgap of dispersion relation (\ref{linear}), i.e.,
at
\begin{equation}
\mu =\mp \left( \Omega -\delta \mu \right) ,0<\delta \mu \ll \Omega ~,
\label{close}
\end{equation}
the two-component problem can be reduced to one of those previously solved
for the single component in the semi-infinite gap \cite{Pedri,Tikhonenkov}.
Close to the bottom (top) edge, Eq. (\ref{Basiceq2}) for $u_{-}$ ($u_{+}$)
makes it possible to eliminate this component in favor of $u_{+}$ ($u_{-}$):
\begin{equation}
u_{\mp }\approx \left( \lambda /2\Omega \right) \left( {\partial }_{x}\pm
i\partial _{y}\right) u_{\pm }.  \label{un_bottom}
\end{equation}%
In particular, for the isotropic semi-vortex represented by ansatz (\ref{SV}%
), Eq. (\ref{un_bottom}) amounts to $U_{-}(r)\approx \left( \lambda /2\Omega
\right) dU_{+}/dr$.

Substituting relation (\ref{un_bottom}) in Eq. (\ref{Basiceq2}) for $u_{\pm
} $, one arrives at an equation for the single component:
\begin{equation}
\delta \mu \cdot u_{\pm }={\frac{\lambda ^{2}}{2\Omega }}\nabla ^{2}u_{\pm
}\pm g|u_{\pm }|^{2}u_{\pm }\pm \kappa u_{\pm }(\mathbf{r})\int R(\mathbf{r}-%
\mathbf{r^{\prime }})|u_{\pm }(\mathbf{r^{\prime }})|^{2}d\mathbf{r^{\prime }%
},  \label{isotropic}
\end{equation}%
where $\nabla ^{2}=(\partial _{x}+i\partial _{y})(\partial _{x}-i\partial
_{y})\equiv \partial _{x}^{2}+\partial _{y}^{2}$ appears as a square of the
SOC operator from Eqs. (\ref{Basiceq01}) and (\ref{Basiceq02}). It is easy
to see that the same $\nabla ^{2}$ is produced by squaring the SOC operator
which presents a general combination of the Rashba and Dresselhaus terms.

Along with Eq. (\ref{isotropic}), it is relevant to consider its
time-dependent version,
\begin{equation}
\mp i\frac{\partial }{\partial t}\tilde{u}_{\pm }={\frac{\lambda ^{2}}{%
2\Omega }}\nabla _{\pm }^{2}\tilde{u}\pm g|\tilde{u}_{\pm }|_{\pm }^{2}%
\tilde{u}\pm \kappa \tilde{u}_{\pm }(\mathbf{r})\int R(\mathbf{r}-\mathbf{%
r^{\prime }})|\tilde{u}_{\pm }(\mathbf{r^{\prime }})|^{2}d\mathbf{r^{\prime }%
},  \label{time}
\end{equation}%
where $\tilde{u}_{\pm }\left( x,y,t\right) \equiv \exp \left( \pm i\Omega
t\right) ~u\left( x,y,t\right) $. In particular, Eq. (\ref{time}) is used
below to test stability of various solitons generated by Eq. (\ref{isotropic}%
).

{If the contact nonlinear term }$\sim g${\ }is present in Eq. (\ref%
{isotropic}), while the DDI is absent ($\kappa =0$), this equation produces
no solitons in the case of the effective self-defocusing, $\pm g<0$ (recall
we define $\Omega $ to be positive). In the case of the local self-focusing,

$\pm g>0$, Eq. (\ref{isotropic}) produces unstable \textit{Townes solitons}
\cite{Townes}. This argument suggests that the contact interaction of either
sign, in the absence of the DDI, cannot support stable solitons in the
present system. As mentioned above, this expectation is fully corroborated
by numerical results (not shown here in detail).

Proceeding to the opposite case, when the DDI is present, while the contact
interaction is absent ($g=0$), we note that, near the bottom edge of the
bandgap, Eq. (\ref{isotropic}) for $u_{+}$, with $\kappa >0$ and the
isotropic kernel, $R=R_{\mathrm{iso}}(\mathbf{r}-\mathbf{r^{\prime }})$,\
gives rise to the ground state in the form an axisymmetric bright soliton
with zero vorticity \cite{Pedri}, while the respective smaller component $%
u_{-}$, produced\ by Eq. (\ref{un_bottom}), features vortical structure $%
\sim re^{i\theta }$ (at small $r$), conforming to the semi-vortex structure
of the isotropic GS, as given by Eq. (\ref{SV}). In fact, in this case, the
2D integration in Eq. (\ref{isotropic}) can be reduced to the radial-only
integration, with kernel $R_{\mathrm{iso}}(\mathbf{r}-\mathbf{r^{\prime }})$
substituted by the effective radial one, as done above, in the general form,
in Eq. (\ref{rho}).

On the other hand, near the top edge of the bandgap, Eq. (\ref{isotropic})
for $u_{-}$, with $\kappa >0$ and the anisotropic kernel taken as per Eq. (%
\ref{aniso}), gives rise to the ground state in the form of the 2D \textit{%
anisotropic} bright soliton, as previously demonstrated in Ref. \cite%
{Tikhonenkov}, while Eq. (\ref{un_bottom}) produces a smaller component, $%
u_{+}$, in the form of an \textit{anisotropic vortex. }Overall, the
anisotropic GS constructed in this form seems as a deformed semi-vortex, as
corroborated by numerically exact results displayed below in Fig. \ref%
{solitonsolution}(b1,b2,b3).

\subsection{Numerical findings for generic two-component solitons}

The above results predict that the isotropic DDI for dipoles polarized
perpendicular to the $\left( x,y\right) $ plane, and the anisotropic DDI for
the in-plane polarization, support, severally, stable isotropic GSs [with
the semi-vortex structure, as per Eq. (\ref{SV})] and anisotropic 2D GSs,
with $\mu $ taken, respectively, close to the bottom or top edge of the
bandgap. These predictions have been corroborated by numerical solutions of
Eqs. (\ref{Basiceq1}) and (\ref{Basiceq2}). The numerical results are
collected in Fig. \ref{Spectra}(b1,b2) for the system which does not include
the contact interaction [$g,\tilde{g}=0$ in Eqs. (\ref{Basiceq1}) and (\ref%
{Basiceq2})]. The results clearly demonstrate that the quasi-analytical
approximations remain valid not only close to the edges, but actually across
the entire bandgap, and extend, as ESs (see further details below), into the
bands.

\begin{figure}[tbp]
\centering{\label{fig3a} \includegraphics[scale=0.43]{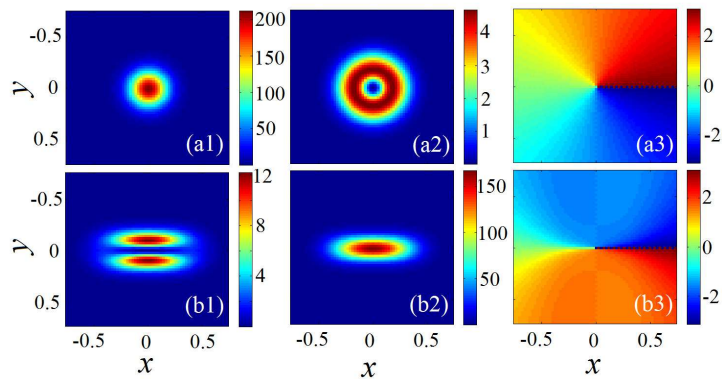}}
\caption{(a1,a2) Density patterns of $u_{+}$ and $u_{-}$ (zero-vorticity and
vortex components, respectively) for a stable isotropic GS (semi-vortex)
with $N=21.37$, $\protect\mu =1.17$. (a3) The phase structure of $u_{-}$.
(b1-b3) The same for a stable anisotropic GS (deformed semi-vortex) with $%
N=22.02$, $\protect\mu =0.016$).}
\label{solitonsolution}
\end{figure}

\begin{figure}[tbp]
\centering{\label{fig4a} \includegraphics[scale=0.3]{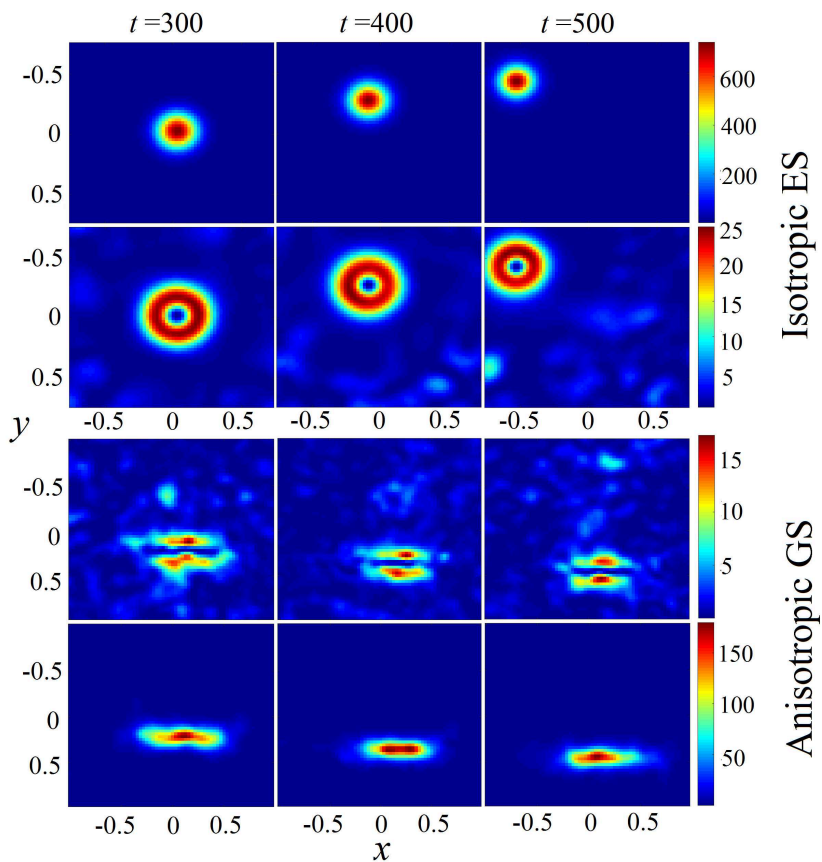}} \centering{%
\label{fig5a}}
\caption{Evolution of an unstable isotropic ES, with $N=56.43$, $\protect\mu %
=24.84$ (a), and of an anisotropic GS with $N=34.5$,~$\protect\mu =-8$ (b).
Density snapshots of the zero-vorticity and vortex components are shown in
the top and bottom panels, respectively, for the isotropic soliton, and vice
versa for the anisotropic one.}
\label{unstable_ES}
\end{figure}

Numerical solutions of Eq. (\ref{Basiceq2}) were produced by means of the
squared-operator method \cite{Jianke2007}. The scaling invariances of Eqs. (%
\ref{Basiceq01}) and (\ref{Basiceq2}) were used to fix $\Omega =10$, $%
\lambda =1$, and $\kappa =0.1$. Generic results were produced fixing the
regularization parameter as $\epsilon =0.5$ (with other reasonable values of
$\epsilon $, similar results have been obtained), while the total norm, $N$,
was varied as an essential control parameter. The stability of the GS
families was identified by means of systematic simulations of the perturbed
evolution (the distinction between stable and unstable states could be
easily detected, as numerical truncation errors were sufficient to trigger
the growth of the instability, if any). As concerns the asymptotic equation (%
\ref{isotropic}), its solutions of were produced by applying the
imaginary-time method \cite{ITM} to its time-dependent version given by Eq. (%
\ref{time}). Comparison of shapes of stable solitons, as found from Eqs. (%
\ref{Basiceq1}) and (\ref{Basiceq2}), and, on the other hand, from the
simplified equations (\ref{isotropic}) and (\ref{time}), is displayed in
Fig. \ref{comparison}.

Branches of isotropic and anisotropic solitons are characterized by $\mu (N)$
dependences displayed in Fig. \ref{Spectra}(b1,b2), along with the
semi-analytical counterparts of these dependences. It has been thus found
that the stable branch of the isotropic GSs extends, across the full
bandgap, into the upper Bloch band abutting on the bandgap, as a family of
ESs, which may exist, under certain conditions, in spectral bands \cite%
{Jianke1999}. In particular, a model supporting ESs in a 2D system was
reported in Ref. \cite{2D-ES}. With the increase of $N$, the isotropic-ES
branch loses its stability inside of the Bloch band, at $N_{\max }\approx 45$
($\mu \approx 16.7$). At $N>N_{\max }$, the branch extends indefinitely into
the band in an unstable form. On the contrary, the stability of the
anisotropic GSs terminates still in the bandgap, at $N_{\max }\approx 28.2$ (%
$\mu \approx $ $-3.8$), the ES continuation of this branch being fully
unstable. The robustness of the solitons in the present system is further
attested to by the fact that unstable ones, both isotropic and anisotropic,
do not suffer destruction, their vortex component keeping its vorticity: as
shown in Fig. \ref{unstable_ES}, unstable solitons commence spontaneous
motion, instead of destruction, emitting small amounts of radiation from the
vortex component. The evolution of the weakly unstable isotropic ESs (recall
all the isotropic GS are stable) does not break their circular symmetry
either.

As said above, isotropic GSs are built as semi-vortices [defined as per Eq. (%
\ref{SV})], i.e., bound states of zero-vorticity and vortex components, as
can be clearly seen in Figs. \ref{solitonsolution}(a1-a3). The extension of
the GS solutions in the ES form keeps the semi-vortex shape as well.
Although no exact ansatz for a vortex structure is available for anisotropic
GSs, Figs. \ref{solitonsolution}(b1-b3) clearly demonstrate that the
anisotropic GSs (and their ES extension) feature the shape of deformed
semi-vortices. An essential peculiarity of the semi-vortices is that their
vortex component carries a relatively small share of the total norm, in
comparison with the zero-vorticity counterpart, as shown in Fig. \ref%
{Spectra}(c) for the isotropic and anisotropic semi-vortices alike.

The stable isotropic solitons shown above, with vorticities $%
(S_{+},S_{-})=(0,1)$ in their large and small components, are fundamental
states, as the system cannot produce any state with a simpler structure.
However, it is possible to look for more complex modes (excited states). To
generate them near the bottom edge of the bandgap, one can use, as a seed,
isotropic vortex-solitons solutions of Eq. (\ref{isotropic}) for the large
component, $u_{+}$, with vorticities $S_{+}=\pm 1$, which are known from the
study of the single-component dipolar BEC \cite{Tikhon2}. Then, Eq. (\ref%
{un_bottom}) generates vorticity $S_{-}=S_{+}+1$ in the small $u_{-}$
component. We have found that both species of the resulting composite modes,
with $(S_{+},S_{-})=(1,2)$ and $(-1,0)$, are unstable against splitting into
a pair of fragments (not shown here in detail). In fact, splitting is a
common instability mode of vortex solitons \cite%
{old-review,Dumitru,recent-review,Skryabin}, although nonlocality may
stabilize some of them \cite{Kiev,Tikhon2}.

\begin{figure}[tbp]
\centering{\label{fig6a} \includegraphics[scale=0.35]{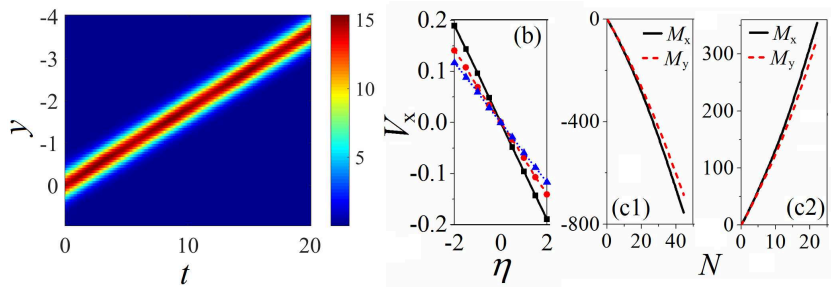}}
\caption{(a) Stable motion of an isotropic GS (semi-vortex) with $N=5.1$ and
$\protect\mu =-8.41$, which was kicked in the $y$-direction with strength $%
\protect\eta =+3$, as per initial conditions (\protect\ref{input}). (b)
Velocity $V_{x}$ of the isotropic GS, kicked along $x$, versus $\protect\eta
$, for $N=2.01$ (black squares), $21.36$ (red circles) and $44.47$ (blue
triangles), respectively. (c1,c2) Negative and positive effective masses, $%
M_{x}$ and $M_{y}$, for isotropic and anisotropic solitons, respectively.}
\label{mobility}
\end{figure}

\subsection{Physical parameters for the gap solitons in the Bose-Einstein
condensate}

It is relevant to estimate actual parameters of the BEC\ solitons which can
be created according to the results reported above in the scaled form. To
this end, we translate the results into physical units corresponding to the
experimental realization of SOC \cite{Zeeman,Cr,Xunda,Drummond-Pu}, taking
values of the magnetic moment for $^{52}$Cr or $^{164}$Dy atoms, and the
strength of the transverse trapping potential $\omega _{\perp }\sim 100$ Hz.
We thus conclude that the stable quasi-2D solitons may be created with the
number of atoms in the range from $N_{\mathrm{at}}\sim 10^{3}$ (near edges
of the bandgap) $\ $to $N_{\mathrm{at}}\sim 10^{4}$ (deeper in the bandgap),
and physical lateral sizes $l_{\mathrm{phys}}\sim 10$ $\mathrm{\mu }$m. The
corresponding relations between the physical quantities and scaled ones
displayed in Figs. \ref{Spectra}-\ref{consider_g} is%
\begin{equation}
N_{\mathrm{at}}\sim 10^{3}N,~\left( x,y\right) _{\mathrm{phys}}\sim
(x,y)\times 20\mathrm{\mu m~.}  \label{Nat}
\end{equation}%
Further, the magnetic field necessary for inducing the appropriate ZS is
estimated as $H\sim 0.1-1$ G. In addition to the magnetic realization, the
use of a spinor BEC\ built of small molecules carrying electric dipole
moments \cite{electric} may be feasible too. Another possibility for the
realization of the stable 2D solitons predicted above is to use the DDI
between moments induced by an external dc magnetic or electric field \cite%
{induced}.

\section{Mobility and collisions of the isotropic and anisotropic solitons}

Soliton mobility in the system under the consideration is a nontrivial
issue, as Eqs. (\ref{Basiceq01}) and (\ref{Basiceq02}) are not Galilean
invariant, although they conserve the total momentum,
\begin{equation}
\mathbf{P}=i\int \left[ \left( \nabla \phi _{+}^{\ast }\right) \phi
_{+}+\left( \nabla \phi _{-}^{\ast }\right) \phi _{-}\right] d\mathbf{r},
\label{P}
\end{equation}%
and their asymptotic version, which amounts to the single equation (\ref%
{time}), remains Galilean invariant. The mobility was tested, in the
framework of the system without the contact interactions ($g=\tilde{g}=0$),
by applying kick $\eta $ to stable quiescent solitons in the $x$ or $y$
direction, which correspond to simulating Eqs. (\ref{Basiceq01}) and (\ref%
{Basiceq02}) with input%
\begin{equation}
\phi _{\pm }(x,y,t=0)=u_{\pm }(x,y)\left\{ e^{i\eta x},e^{i\eta y}\right\} ,
\label{input}
\end{equation}%
the components of the respective momentum (\ref{P}) being $P_{x,y}=N\eta $.
An example of stable motion of a kicked isotropic GS (semi-vortex) is shown
in Fig. \ref{mobility}(a). The velocity of the moving isotropic soliton, $%
V_{x}$, is displayed in Fig. \ref{mobility}(b) as a function of the kick's
magnitude, $\eta $. It is seen that the mass of kicked isotropic GSs is
\emph{negative}, as they move in the direction \emph{opposite} to the kick.
It is relevant to mention that the negative dynamical mass is a generic
feature of GSs known in other settings, including 2D GSs \cite{HS,plasmon}.

The effective mass of the isotropic solitons in the $x$ and $y$ directions,
defined as $M_{x,y}(N)=P_{x.y}/V_{x,y}$, is displayed in Fig. \ref{mobility}%
(c1) as a function of $N$ (strictly speaking, the motion makes the soliton
slightly anisotropic). Naturally, the mass increases almost linearly with
the norm. Close to the bottom edge of the bandgap, where $N$ is very small,
the mass is isotropic, $M_{x}\approx M_{y}$, as the corresponding equation (%
\ref{isotropic}) is isotropic too. Farther from the bandgap edge, the mass
features a weak anisotropy. On the other hand, the anisotropic solitons
feature a positive mass, which also grows almost linearly with $N$, as shown
in Fig. \ref{mobility}(c2). In the course of its motion, the kicked solitons
show very weak transverse displacement, due to the \textquotedblleft Magnus
force" acting on the small vortical component \cite{Magnus}. Additional
high-accuracy simulations are needed to study the latter effect in detail.

\begin{figure}[tbp]
\centering{%\label{fig5a}
\includegraphics[scale=0.45]{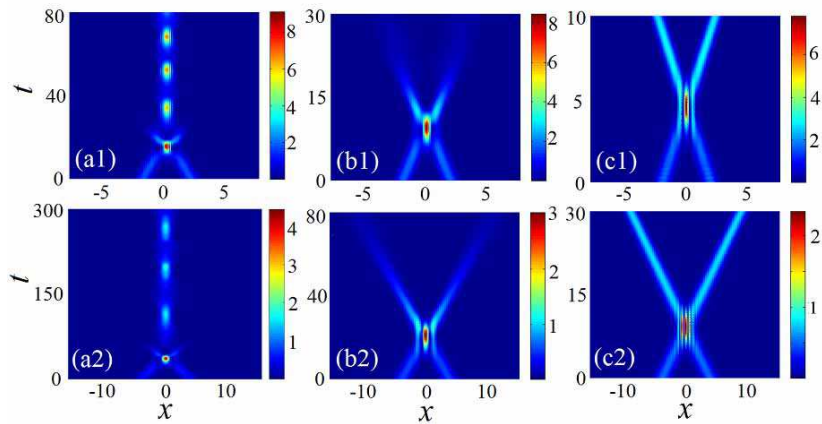}}
\caption{(a1-c1): Collisions between stable isotropic GSs with $N=2.013,%
\protect\mu =-9.7$. The solitons were set in motion by opposite kicks: $%
\protect\eta =\pm 1$ (a1), $\pm 2$ (b1), $\pm 5$ (c1). (a2-c2): The same for
anisotropic GSs, with $N=2.004,\protect\mu =9.94$, and kicks $\protect\eta %
=\pm 1$ (a2), $\pm 2$ (b2), $\pm 5$ (c2).}
\label{collision}
\end{figure}

It is natural too to study collisions between stable GSs set in motion by
opposite kicks $\pm \eta $. The simulations reveal three generic outcomes of
the collisions, for the isotropic and anisotropic solitons alike: merger
into a single breather [Figs. \ref{collision}(a1,a2)], elastic collision
[Figs. \ref{collision}(c1,c2)], and transformation into diffracting
quasi-linear beams [Figs. \ref{collision}(b1,b2)], at, small, large, and
intermediate values of $\eta $, respectively.

%\begin{figure}[tbp]
%\centering{%\label{fig5a}
%\includegraphics[scale=0.35]{5a.eps}}
%\caption{Collision were simulated for pairs of GSs taken with a sufficiently large
%separation $2x_{0}$ and opposite kicks applied to them, i.e., with initial
%conditions $\phi^{(t=0)} _{\pm }=u_{\pm }(x+x_{0},y)e^{i\eta x}+u_{\pm}(x-x_{0},y)e^{-i\eta x}$.
%For negative-mass isotropic GSs, the collision occurs with $\eta <0$, while for positive-mass anisotropic GSs, %the collision occurs with $\eta >0$. (a1-a3) Collisions between stable isotropic GSs with $N=2.013,%
%\protect\mu =-9.7$. The solitons were set in
%motion by opposite kicks: $\eta =-1$ (a1), $-2$ (b1), and $-
%5$ (c1). (a2-c2). The same for anisotropic GSs with $N=2.004,\protect\mu %
%=9.94$, and $\protect\eta = 1$ (a2), $2$ (b2), and $5$ (c2).}
%\label{collision}
%\end{figure}

\section{Effects of the contact interaction}

The local mean-field nonlinearity, induced by interatomic collisions, is
always present in the bosonic gas, therefore it is relevant to explore
effects of the contact interaction on the soliton families obtained above in
the absence of the contact terms in Eqs. (\ref{Basiceq1}) and (\ref{Basiceq2}%
). Here, we perform this analysis in the framework of the asymptotic
equations (\ref{isotropic}) and (\ref{time}), as they are sufficient to
capture main effects produced by the local nonlinearity, as shown below.

{First, Eq. (\ref{isotropic}), that includes the contact-interaction term }$%
\sim g${, implies that, due to the possibility of the \textit{critical
collapse} \cite{Townes} in the same 2D equation, the norm of wave function }$%
u_{\pm }$ {cannot exceed a critical value, which is determined by the scaled
norm of the Townes' soliton, }$N_{\mathrm{Townes}}\approx 5.85$ {\cite%
{Townes}, \textit{viz}., }${N_{\pm }<N_{\mathrm{Townes}}/}\left( \pm g\Omega
\right) $. In other words, for given norm $N$, isotropic and anisotropic
self-trapped modes exist, respectively, at%
\begin{equation}
{g<g_{\mathrm{Townes}}^{\mathrm{(iso)}}={N_{\mathrm{Townes}}}/}\left( {%
\Omega N}\right) ,~{g>g_{\mathrm{Townes}}^{\mathrm{(aniso)}}=-{N_{\mathrm{%
Townes}}}/}\left( {\Omega N}\right)  \label{boundary}
\end{equation}%
(recall $\Omega =10$ is fixed in this paper).

\begin{figure}[tbp]
\centering{\ \includegraphics[scale=0.22]{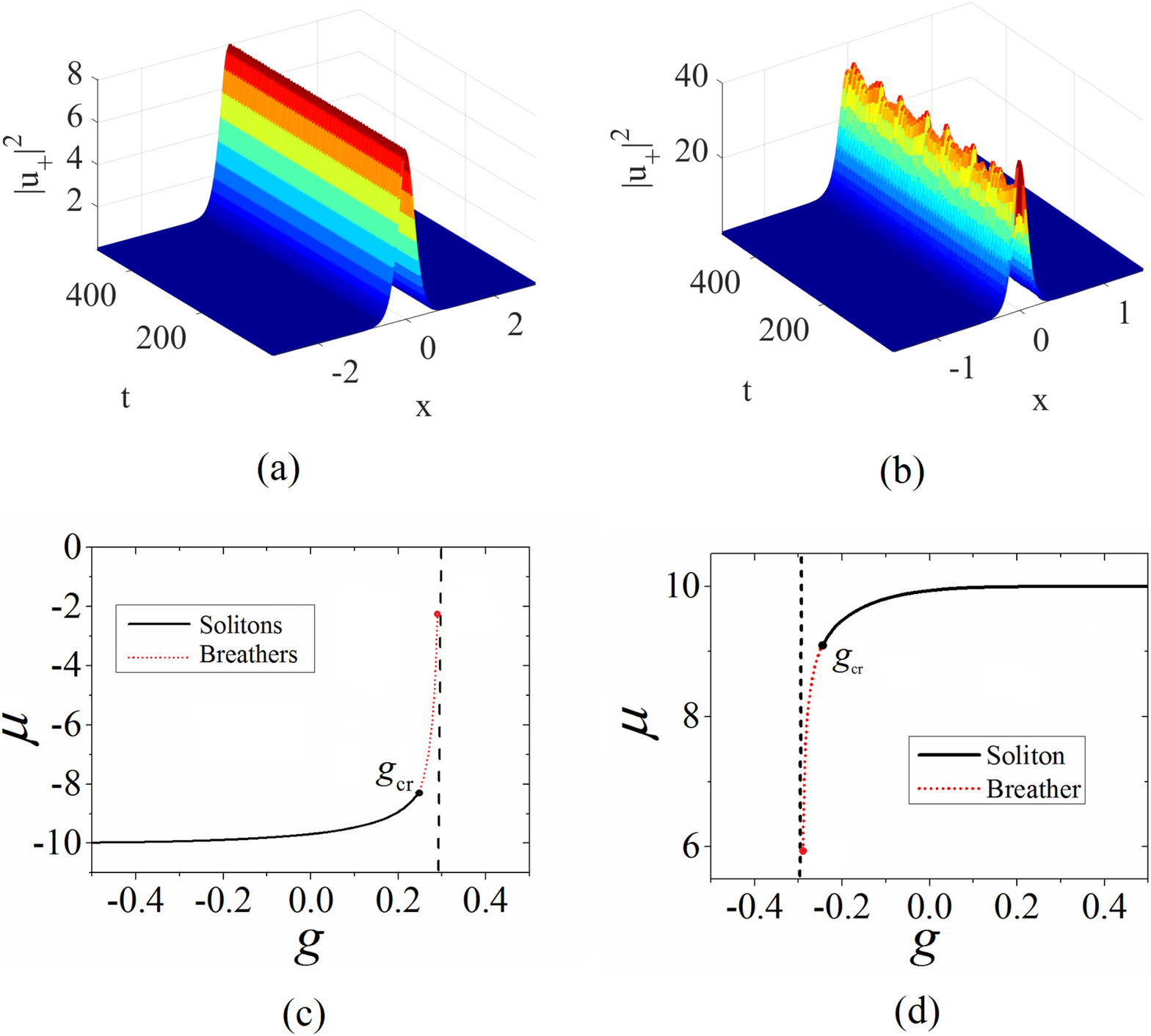}}
\caption{(a,b) Simulations of the evolution of isotropic GSs, in the
framework of Eqs. (\protect\ref{Basiceq01}) and (\protect\ref{Basiceq02})
which include the local nonlinear terms with $g=0.2$ (a) and $g=0.28$ (b),
while the norm of the large zero-vorticity component is fixed as $N_{+}=2$.
The results are displayed for the cross section of $\left\vert \protect\phi %
_{+}(x,y,t)\right\vert ^{2}$ at $y=0$. (c) The chemical potential of
isotropic solitons with $N_{+}=2$ versus $g$. Black solid and red dotted
segments represent, severally, stable stationary solitons and persistent
breathers, which replace unstable solitons at $g>g_{\mathrm{cr}}\approx 0.25$%
. The vertical dashed line is the existence boundary for self-trapped modes,
where the collapse sets in, $g=g_{\mathrm{Townes}}^{\mathrm{(iso)}}\approx
0.29$, see Eq. (\protect\ref{boundary}). (d) The chemical potential of the
anisotropic solitons with $N_{-}=2$ versus $g$. The meaning of the black
solid and red dotted segments is the same as as in (c), the boundary between
them being $g_{\mathrm{cr}}\approx -0.25$, while the existence/collapse
boundary is marked by the vertical dashed line at $g=g_{\mathrm{Townes}}^{%
\mathrm{(aniso)}}\approx -0.29$. }
\label{consider_g}
\end{figure}

Furthermore, inside the existence regions (\ref{boundary}), systematic
simulations of Eq. (\ref{time}) have revealed an intrinsic stability
boundary, $g=${\ $g_{\mathrm{cr}}^{(\pm )}\approx \pm 0.25$, such that the
isotropic and anisotropic stationary GSs are stable, respectively, at }$g<${%
\ $g_{\mathrm{cr}}^{(+)}$ and $g>g_{\mathrm{cr}}^{(-)}$, while in the
remaining intervals, $g_{\mathrm{cr}}^{(+)}<g<$}${g_{\mathrm{Townes}}^{%
\mathrm{(iso)}}}$ and ${g_{\mathrm{Townes}}^{\mathrm{(aniso)}}<g<}${$g_{%
\mathrm{cr}}^{(-)}$, the GSs are unstable, spontaneously transforming into
persistent breathers, as shown in Fig. \ref{consider_g}. Thus, we conclude
that the stationary GSs exist and remain completely stable when the \emph{%
arbitrarily strong} contact interaction is\emph{\ self-repulsive}, in terms
of Eqs. (\ref{isotropic}) and (\ref{time}) (which corresponds to }$g<0$ and $%
g>0$ for the isotropic and anisotropic GSs, respectively){, being
compensated by the effectively attractive DDI. In the limit of very strong
local self-attraction, the solitons become very broad, corresponding to }$%
\delta \mu \rightarrow 0$ in terms of Eq. (\ref{isotropic}), i.e., $\mu
\rightarrow -\Omega $ and $\mu \rightarrow +\Omega $, in terms of Figs. {\ref%
{consider_g}(c) and (d), respectively. Note an essential difference of these
limits from those shown in Figs. \ref{Spectra}(b1) and (b2): in the latter
case, the soliton's norm vanishes at the edges of the bandgap, while in the
cases displayed in }Figs. {\ref{consider_g}(c) and (d) the soliton branches
keep the fixed finite norm, }$N_{\pm }=2$.

To explicitly compare strengths of the competing contact interactions and
DDI, we define the relative strength,
\begin{equation}
\epsilon _{\mathrm{dd}}\equiv g_{\mathrm{eff}}^{(%
\mathrm{iso,aniso})}/g,  \label{effective_epsilon}
\end{equation}%
where $g_{\mathrm{dd}}^{(\mathrm{iso,aniso})}$ are effective DDI
coefficients for the isotropic and the anisotropic GSs, respectively, which
are defined, for the definiteness' sake, at centers of the solitons, as
follows:
\begin{equation}
\kappa u_{+}(\mathbf{r}=0)\int R_{\mathrm{iso}}(0-\mathbf{r^{\prime }}%
)|u_{+}(\mathbf{r^{\prime }})|^{2}d\mathbf{r^{\prime }\equiv }g_{\mathrm{eff}%
}^{(\mathrm{iso})}|u_{+}(\mathbf{r}=0)|^{2}u_{+}(\mathrm{r}=0),  \label{int+}
\end{equation}%
\begin{equation}
\kappa u_{-}(\mathbf{r}=0)\int R_{\mathrm{aniso}}(0-\mathbf{r^{\prime }}%
)|u_{-}(\mathbf{r^{\prime }})|^{2}d\mathbf{r^{\prime }\equiv }g_{\mathrm{eff}%
}^{(\mathrm{aniso})}|u_{-}(\mathbf{r}=0)|^{2}u_{-}(\mathrm{r}=0).
\label{int2}
\end{equation}%
Here, we adopt conditions $|u_{+}(\mathbf{r})|^{2}\gg |u_{-}(\mathbf{r})|^{2}
$ and $|u_{-}(\mathbf{r})|^{2}\gg |u_{+}(\mathbf{r})|^{2}$ for the isotropic
and anisotropic GSs, respectively, as Figs. \ref{consider_g}(c) and (d)
clearly demonstrates that these conditions hold at the critical (most
essential) points.

\begin{figure}[tbp]
\centering{\ \includegraphics[scale=0.35]{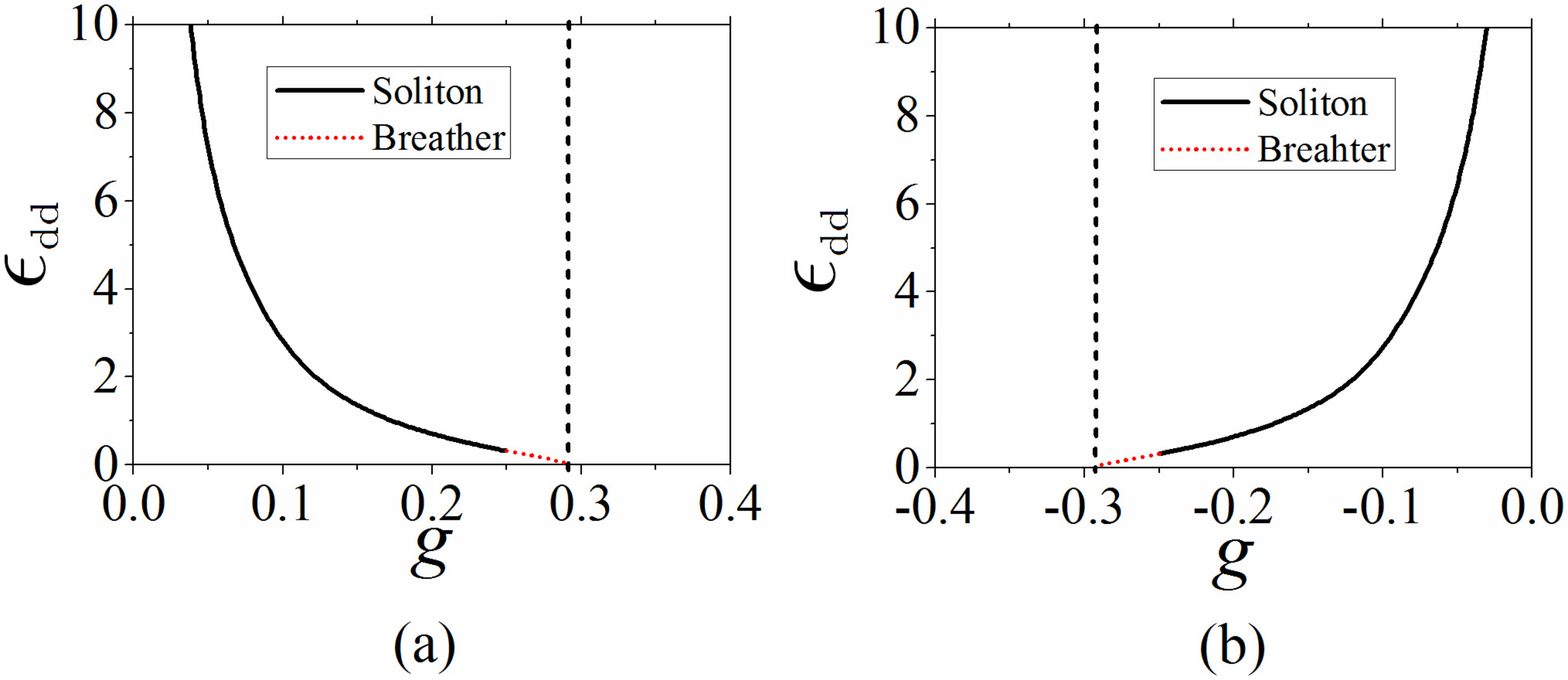}}
\caption{The relative strengths of the DDI, $\epsilon_{\mathrm{dd}}$, defined as per
Eqs. (\ref{effective_epsilon})-(\ref{int2}), are displayed as functions
of $g$. Black solid and red dotted segments, as well as the vertical dashed lines,
have the same meaning as in Fig. \ref{consider_g}.
(a) The $\epsilon_{\mathrm{dd}}(g)$ dependence for isotropic solitons with
$N_{+}=2$. (b) The same for anisotropic solitons with $%
N_{-}=2$.}
\label{epsilon_g}
\end{figure}

The so defined relative DDI strengths (\ref{effective_epsilon}) are
displayed, as functions of $g$, in Fig. \ref{epsilon_g}, for isotropic and
anisotropic GSs at $g>0$ and $g<0$, respectively. A natural conclusion is
that the GSs suffer the destabilization when the DDI becomes too weak in
comparison with the contact interaction.

On the contrary to the setting with the competing contact interaction and
DDI, the interplay of the local and nonlocal nonlinear interactions with
identical signs leads to the destabilization of the solitons in the present
system, as is shown by Figs. \ref{consider_g}(c,d). Overall, the present
situation qualitatively resembles that reported in Ref. \cite{Minhang}, in
which a binary BEC represented a mixture of two atomic states coupled by a
microwave field. In such a setting, nonlocal attraction between the
components, mediated by the microwave field, was sufficient for the
existence and stability of two-component solitons in the presence of
arbitrarily strong local self- and cross- repulsion.

\section{Conclusion}

The objective of this work is to propose a setting for the creation of
stable quasi-2D GSs (gap solitons) and ESs (embedded solitons) in free
space. The system is based on the spinor dipolar BEC, whose components are
coupled by the SOC (spin-orbit interaction). For quasi-2D states, the
kinetic-energy terms in the spinor GPEs are negligible in comparison with
SOC, which gives rise to simplified couple-mode equations, with the bandgap
provided by the ZS (Zeeman splitting). Stable isotropic and anisotropic 2D\
solitons were thus found, in the quasi-analytical and numerical forms, for
the dipoles polarized perpendicular and parallel to the system's plane,
respectively. Both families continue as ESs into adjacent spectral bands,
the isotropic-ES branch being partly stable. Mobility and collisions of the
solitons were studied too, concluding that the mass of the
isotropic/anisotropic ones is negative/positive. Effects of contact
interactions, added to the DDI (dipole-dipole interaction), were studied
too, with a conclusion that the stationary GSs persist and remain stable in
the presence of the arbitrarily strong local self- and cross-component
repulsion, compensated by the effectively attractive DDI.

A challenging possibility is to extend the present analysis from solitons to
``quantum droplets" in binary dipolar BEC, in the presence of SOC.
``Droplets" stabilized by beyond-the-mean-field effects were recently
created in single-component dipolar condensates \cite{droplets}.

\begin{acknowledgments}
This work was supported, in part, by NNSFC (China) through Grant No.
11575063, and by the joint program in physics between NSF and Binational
(US-Israel) Science Foundation through project No. 2015616. B.A.M.
appreciates a foreign-expert grant of the Guangdong province (China).
\end{acknowledgments}

\bibliographystyle{plain}
\bibliography{apssamp}
% Produces the bibliography via BibTeX.

\end{document}